\documentclass[a4paper]{jpconf}
\usepackage{graphicx}
\usepackage{amsmath}
\usepackage{bbold}
\usepackage{epsfig}
\usepackage[latin1]{inputenc}

\def \begineq{\begin{equation}}
\def \endeq{\end{equation}}
\def \({\left(}
\def \){\right)}
\def \[{\left[}
\def \]{\right]}

\begin{document}
\title{Performance of simulated annealing in $p$-spin glasses}

\author{Florent Krzakala$^{\dag,\ddag}$ and Lenka Zdeborov\'a$^\star$}

\address{$\dag$ Laboratoire de Physique Statistique, CNRS UMR 8550,
  Universit\'e P. et M. Curie, Ecole Normale Sup\'erieure, Paris,
  France.}  \address{$\ddag$ ESPCI and CNRS UMR 7083, 10 rue Vauquelin
  Paris 75005 France} \address{$\star$ Institut de Physique
  Th\'eorique, IPhT, CEA Saclay and URA 2306, CNRS, 91191
  Gif-sur-Yvette, France.}

\ead{florent.krzakala@ens.fr, lenka.zdeborova@cea.fr}

\begin{abstract} 
  We perform careful numerical simulations of slow Monte-Carlo
  annealings in the dense 3-body spin glass model and compare with the
  predictions from different theories: thresholds states,
  isocomplexity, following state. We conclude that while isocomplexity
  and following state both provide excellent agreement the numerical
  data, the influence of threshold states -- that is still the most
  commonly considered theory -- can be excluded from our data.
\end{abstract}

Predicting the dynamical behavior from static calculations is the
holy grail of theoreticians in statistical physics. After all, this is
precisely what statistical mechanics is about from the very beginning:
we prefer to consider the static average over all configurations
rather than the complex dynamics of all atoms. A particularly
important case, both in classical and quantum thermodynamics, is given
by the dynamics after very slow variations in an external parameter so
that the system remains at equilibrium. Such changes are said to be
"adiabatic". When a macroscopic system is in a given phase and if one
tunes a parameter, say the temperature, or the external field, very
slowly then all observables, such as the energy or the magnetization
in a magnet, will be given by the equilibrium equation of state.
Given a system at equilibrium in a well-defined phase, it is always
possible to consider the adiabatic evolution: In the low-temperature
phases of a ferromagnet, for instance, the evolution of the
magnetization is different in the two phases or Gibbs states
corresponding to the positive or negative magnetization.

To describe this theoretically, one can force the system to be in the
Gibbs state of choice for instance, by adding an external
infinitesimal field or fixing the boundary conditions and then study
the adiabatic evolution for each of these phases. This simplicity
breaks down in glassy systems: they have a complicated rugged,
many-valleys energy landscape with an enormous number of different
Gibbs states. The statics picture and the statistical features of the
landscape are well known in the mean field case. However, with the
exception of few models
\cite{0305-4470-28-15-003,CugliandoloKurchan93,BouchaudDean95,BouchaudCugliandolo98},
an analytical description of the dynamics, and of the way the Gibbs
states are evolving upon adiabatic changes, has been missing. Note that
by "adiabatic" we mean here a slow dynamics that takes time slower
than any power of the system size, but we want to avoid exponentially
slow change, where of course we stay always at equilibrium but that
has little relevance experimentally if one is not ready to wait a time
longer than the age of the universe.

This is by no means an academic exercice. In fact, it is a fundamental
question one needs to ask in both physics and computer
science. Consider the latter: simulated annealing~\cite{kirkpatrick:83}
is one of the most important contribution from statistical physics to
optimization and computer science. Consider an annealing experiment
where temperature $T$ is changed slowly in time: what is the limiting
energy of such an annealing? If one changes slowly the pressure and
compresses a set of hard spheres, using a mixture to avoid
crystallization, what is the density of the final packing (this is the
jamming problem)? Clearly, it would be nice to answer such questions
without having to solve the dynamics itself, which is a very difficult
and almost intractable problem in general. In fact, since glassy
systems are never in equilibrium, one may in fact argue that getting
the answer of such questions is more important, and more relevant,
than the static solution corresponding to infinite waiting times. 

With this motivation in mind, the present authors have discussed a
formalism to describe the adiabatic evolution of these glassy Gibbs
states as an external parameter is tuned, and applied it to mean field
spin models in
\cite{krzakala2010following,zdeborova2010generalization}. The purpose
of the present contribution is to discuss numerically the validity of this
approach ---and of related ones--- for the well known and studied
fully connected $p$-spin glass
model \cite{GrossMezard84}.

\section{Following state adiabatically}
Let us start by briefly reminding the principle of the state following
approach. It was pioneered by Franz and Parisi when they derived the
"potential" \cite{FranzParisi97,BarratFranz97,FranzParisi98}. It was
then discussed thoroughly in 
\cite{krzakala2010following,zdeborova2010generalization} and some
suggested by \cite{KrzakalaKurchan07} have been
recently constructed in \cite{franz2012quasi}. Here we shall consider
only the simplest version.

How can one follow adiabatically a given Gibbs state? Consider again
the example of the ferromagnet with the two "up" and "down"
equilibrium states. We can force the system to be in the Gibbs state
of choice by fixing the all negative or the all positive boundary
conditions. Even far away from the boundaries, the system will stay in
the selected state for all $T<T_c$ (above the Curie point any boundary
condition will result in a trivial paramagnetic state). By solving the
thermodynamics conditioned to the boundaries, we can thus obtain the
adiabatic evolution of each of the two states.  What boundary
conditions should be applied in glassy systems where the structure of
Gibbs states is very complicated?  The answer is provided by the
following gedanken experiment: consider an equilibrium configuration
of the system at temperature $T_p$. Now freeze the whole system except
a large hole in it. This hole is now a subsystem with a boundary
condition typical for temperature $T_p$. If the system is in a
well-defined state, then no matter the size of the hole, it will
always remain correlated to the boundaries and stay in the same
state. One may now change the temperature and study the adiabatic
evolution of this state\footnote{The experienced reader will have
  recognized the usual construction for the nucleation argument
  adapted for glassy systems as in the "mosaic picture" of
  \cite{MosaicOld,Mosaic}, but generalized with different
  temperature. This construction is also intimately related to the
  Franz-Parisi potential \cite{FranzParisi97}.}.

In mean-field systems, this construction allows for an analytic
treatment in the spirit of the cavity and replica method. We refer the
reader to \cite{krzakala2010following,zdeborova2010generalization}
description of the method and of the equation, and some comparison
with Monte-Carlo simulations for sparse systems. Here, we shall
consider the fully-connected $p$-spin model, where the following state
method was first put forward by Franz and Parisi in the spherical case
\cite{BarratFranz97}, although we will consider the Ising
version. Again, the computation were done in
\cite{krzakala2010following,zdeborova2010generalization}.

For a temperature annealing in a one-step replica symmetry breaking
model such as the $p$-spin, the idea is the following: At high $T$, a
paramagnetic/liquid state exists and therefore a slow annealing should
be able to follow the equilibrium computation. Below the dynamical
glass temperature $T_d$, this state shatters into exponentially many
Gibbs states, all well separated by extensive energetic or entropic
barriers, leading to a breaking of ergodicity and to the divergence of
the equilibration time. The idea of the following state (and of the
isocomplexity computation as well,see \cite{MontanariRicci04}) is thus
to follow the state that appear at $T_d$ as the temperature is
lowered; this should mark the limit of any non-exponentially slow
annealing.

\section{The p-spin model: a standard benchmark, and a few tricks}
The fully-connected 3-spin glass model reads
\cite{GrossMezard84}: 
\begin{equation}
{\mathcal H}=\sum_{ijk} J_{ijk} S_i S_j S_k\, ,
\end{equation}
where the sum is over all possible triplets of spins, and where the
$J_{ijk}$ are quenched random variables with distribution
\begin{equation}
P(J)=\sqrt{\frac{N^2}{\pi p!}} \exp{\[  -\frac{ N^2}{p!} \]}\, .
\end{equation}
This is the mean field spin glass model at the basis of the mean-field
theory of glasses~\cite{kirkpatrick:89}.

Our goal here is to perform careful Monte-Carlo simulations of
annealings for this
densely connected model which as far as we know, and perhaps surprisingly, was not done yet in the literature.  One
sees that simulating this model is quite computationally demanding since
computing the energy takes $N^3$ operations. 

The first trick we will use is to {\it not} use this model, but rather
a diluted version of it where instead of using all the $N^3$ possible
triplets of spin, we will use $N\sqrt{N}$ of them. In fact, this is
like a simulation of the XOR-SAT model (which is nothing else than the
diluted p-spin, that is, the p-spin on random hyper-graphs) with a
connectivity $c=\sqrt{N}$. Both this model and the fully connected one
have exactly the same static solution as $N\to \infty$ (up to
rescaling). For the sizes we will consider, the difference with the
fully connected model is negligible, but this reduced connectivity
allows for efficient simulations. All the present data are made with
$N=50000$ (for annealing) or $N=20000$ (for quenches). These size are
sufficiently large such that the results do not change visibly when
the size is changed by a multiplicative factor two, hence there is no
need for finite size scaling analysis which simplified greatly our
analysis.

The second trick will be to start with initially equilibrated
configuration. That may seem a hard task in such a glassy model,
however, using the planting trick
\cite{OZEKI,MontanariSemerjian06,4691011,krzakala2009hiding,krzakala2011meltingI,krzakala2011meltingII,ZdeborovaK11},
this is something easy to achieve. In a nutshell, the idea of  quiet planting is to
\textit{first} generate a configuration of spins that we want to be an
equilibrium one, and \textit{then} to create the disorder in the
Hamiltonian such that this is precisely an equilibrium
configuration. Of course, an instance of the problem created in such
a way have no reason to be a ``typical'' one; however, as discussed in
the references above, if the problem admit a ``annealed solution''
(that is, if the annealed solution is the correct one, as it is for
the p-spin as long as $T\ge T_K$), then instances created by the
planting tricks are typical ones in the large $N$ limit.  We refer to
the aforementioned publication for details.

\section{An historical perspective}

\begin{figure}[!ht]
\center
\includegraphics[width=0.6\textwidth]{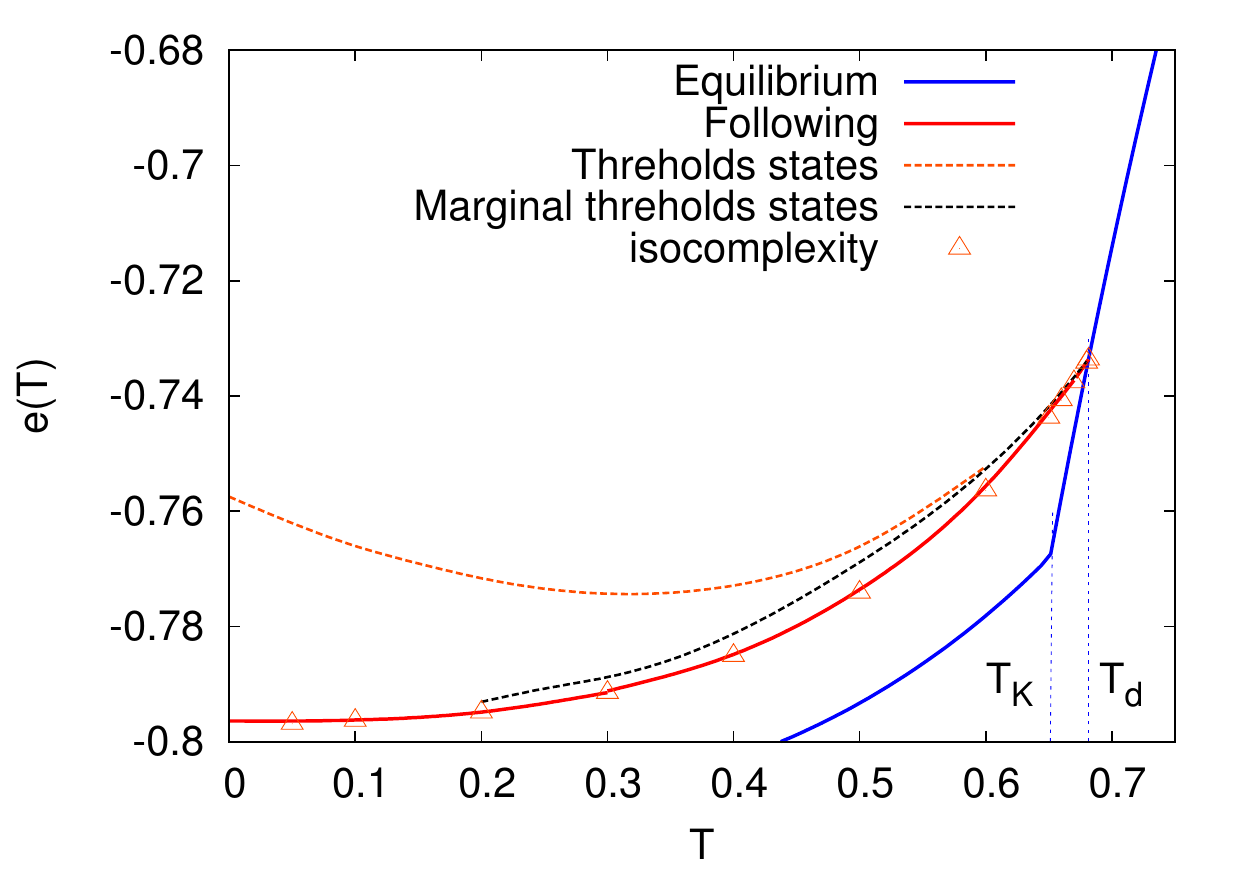}
\caption{\label{fig:0}Results of different computations for the
  annealing in the 3-spin glass problem (see text).}
\end{figure}

Let us remind the different type of predictions for behavior of slow
annealing; they are summarized in Fig.\ref{fig:0}. 

For a long time, the dominating idea has been that one simply has to
compute the so called "thresholds states". There high energy energy
minima in the energy landscape. The idea of such computation is count
how many minima they are at a given energy, and to compute for which
energy they are most numerous. This, according to this idea, allows to
locate at which energy there are minima that can, most likely, trap
the dynamics. The corresponding energy is depicted as ``thresholds
states'' in Fig.~\ref{fig:0}. This idea is popular, mostly because it
works perfectly in the simple spherical $p$-spin
model\cite{0305-4470-28-15-003,CugliandoloKurchan93,BouchaudDean95,BouchaudCugliandolo98},
however, as discussed for instance in \cite{sun2012following}, the
situations is much more complicated already in the mixed spherical
$p$-spin glass model.

It was soon realized \cite{montanari2003nature,MontanariRicci04} that
in fact most of these states are actually the result of a computation
that is not self-consistent (unstable towards more steps of repica
symmetry breaking), and that the threshold states of ``stable'' (or at
least marginally stable) states is much lower. Computing correctly
those turned out to be a complicate task that has been finally
achieved by Rizzo in \cite{rizzo2013replica}. The result is shown in
Fig.~\ref{fig:0} as ``marginal states''. To be precise, we refer as
threshold states the (unstable) 1RSB ones, and to the  marginal states
the (marginal) full RSB ones \cite{rizzo2013replica}. 

However, as discussed in
e.g. \cite{BarratFranz97,FranzParisi98,montanari2003nature,KrzakalaKurchan07,MontanariRicci04,krzakala2010following,zdeborova2010generalization},
it may be a bit naive to assume that a Monte-Carlo annealing will just
end up in the most numerous states, and that was the motivation for
the idea for following states that started at $T_d$. This is why
iso-complexity was introduced: It is proposed to count the number of
equilibrium states at $T_d$, and then to consider the energies at $T <
T_d$ for which the number of states is equal to the one at
$T_d$. Iso-complexity leads indeed to a lower bound on adiabatic
annealings, because in order to end up at lower energies one would
have to be exponentially lucky. Indeed, if one is trap in one of
$e^{N\Sigma_1}$ similar states at an energy $e_1$, and if the total
number of states at  $e_2<e_1$ is $e^{N\Sigma_2}$, with
$Sigma_2<Sigma_1$, it is exponentially unlikly that any of these
correponds to the one at $e_1$.

In fact, the iso-complexity is only an approximation of the following
state formalism (where we explicitly follow states), and explicit
differences can be seen (see for instance \cite{krzakala2010following,zdeborova2010generalization}) however, since the following state computation itself can be rather
  involved (see \cite{franz2012quasi}) and in some cases (for low
  enough temperature) we could only do an approximated computation
  (see
  \cite{krzakala2010following,zdeborova2010generalization,sun2012following}). As
  shown in Fig.~\ref{fig:0}, however, the results of the two approach
  is extremely close in the 3-spin model (see also
  \cite{krzakala2010following,zdeborova2010generalization} for the
  diluted case), at least in the case when one starts from $T=T_d$
  (actually, there are more devitaion for the the states starting from
  $T<T_d$, as seen in
  \cite{krzakala2010following,zdeborova2010generalization}). The
  question is now how these rather different predictions compare with
  simulations.

\section{Annealings follow the following state computation}
\begin{figure}[!ht]
\hspace{-0.2cm}
\includegraphics[width=0.52\textwidth]{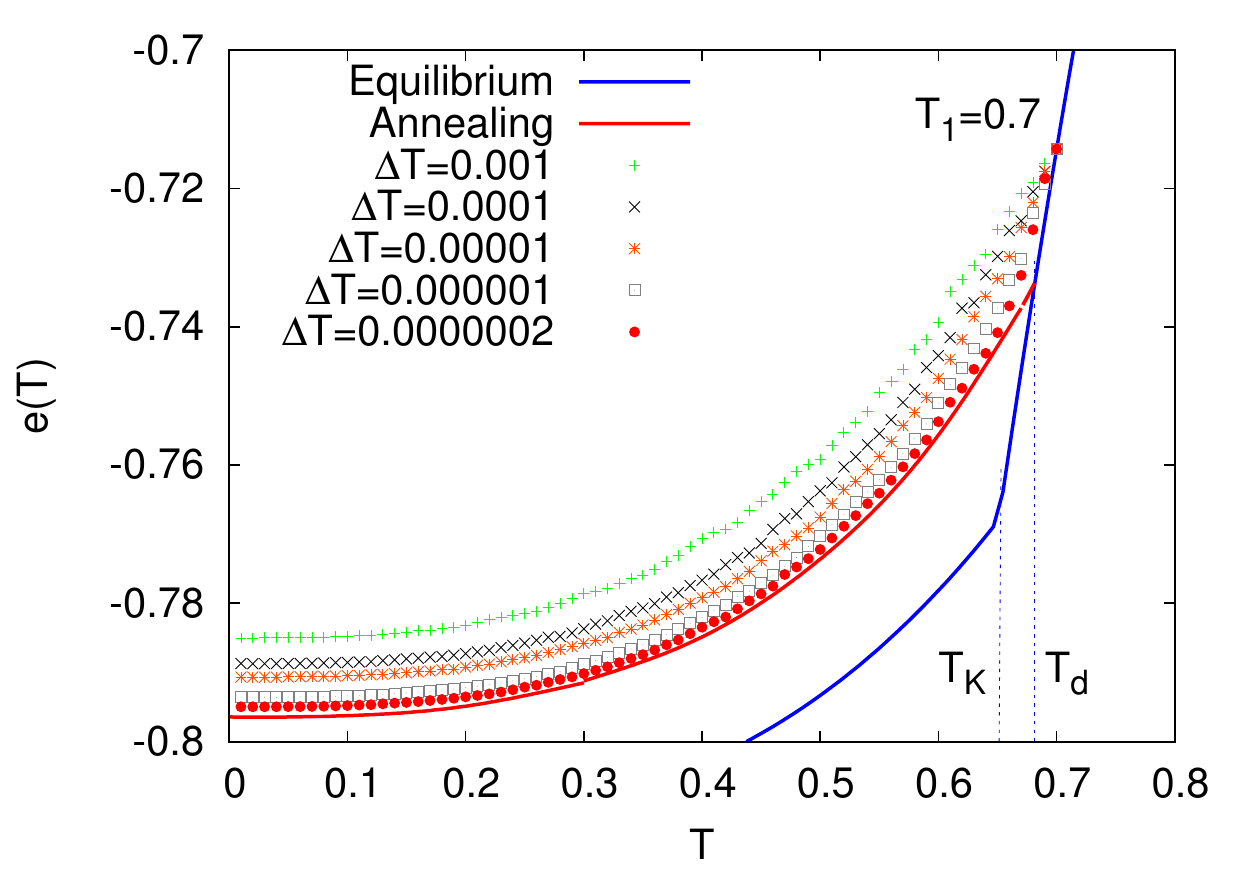}
\hspace{-0.5cm}
\includegraphics[width=0.52\textwidth]{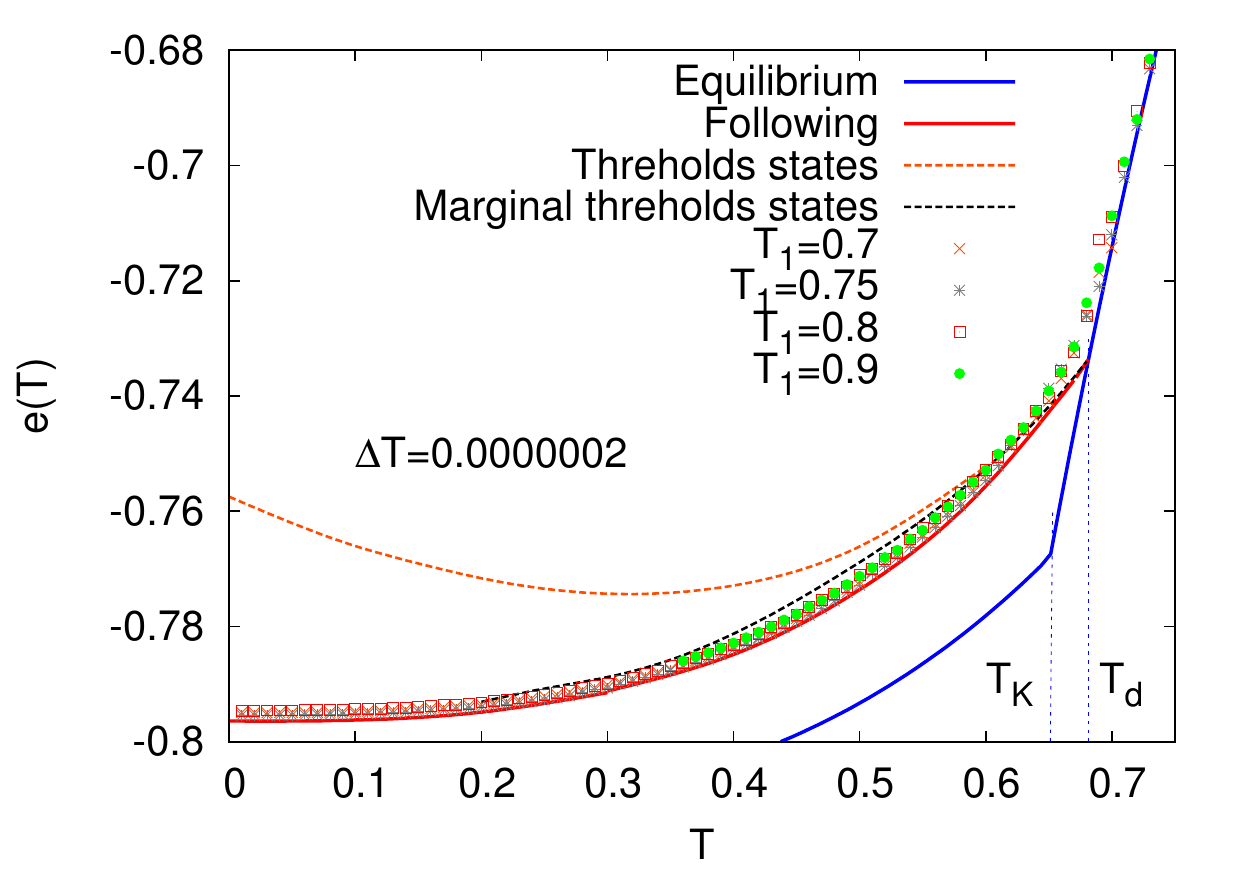}
\caption{\label{fig:1}Annealings starting from an equilibrated
  configuration at high temperature $T_1>T_d=0.6825$ in the 3-spin
  glass problem for $N=50000$. Left: The energy of annealings from
  $T_1=0.7$ with different cooling rates is shown together with the
  equilibrium solution of the model (in blue) and the following state
  prediction (in red). Right: As in the left figure, but with
  different starting points. For comparison, we also shows the energy
  of the thresholds states (dashed orange line) and the marginal
  thresholds states (dashed black line). Annealings have no problems
  going below these energies that do not seem to affect its behavior.}
\end{figure}

The results of annealings starting from an equilibrated configuration
at $T_1>T_d$ are shown in Fig.\ref{fig:1} for different rates. They
are clearly approaching the following state predictions as the rate
decreases $\Delta T \to 0$. We also show these annealing using only
the smallest rate (and demonstrate that the starting temperature does
not matter as long as it is larger than $T_d$). It is striking how
easily the Monte-Carlo dynamics is able to go beyond the thresholds
state, marginal or not.

These data clearly demonstrate that annealings are not {\it at all}
sensitive to the thresholds states, and that the following states of
isocomplexity consideration are the correct ones in this case. It is
also striking how these two are similar (although definitely {\it not}
equal). For models with a discontinuous first order transition, it
thus seem that iso-complexity is a very acceptable approximation. It
would be interesting to see how generic this is (see
\cite{krzakala2010following,zdeborova2010generalization}.  for
discussion in the XORSAT model, \cite{sun2012following} for the
spherical model, and the reviez \cite{ReviewFran} for hard sphere.).

\begin{figure}[!ht]
\hspace{-0.2cm}
\includegraphics[width=0.52\textwidth]{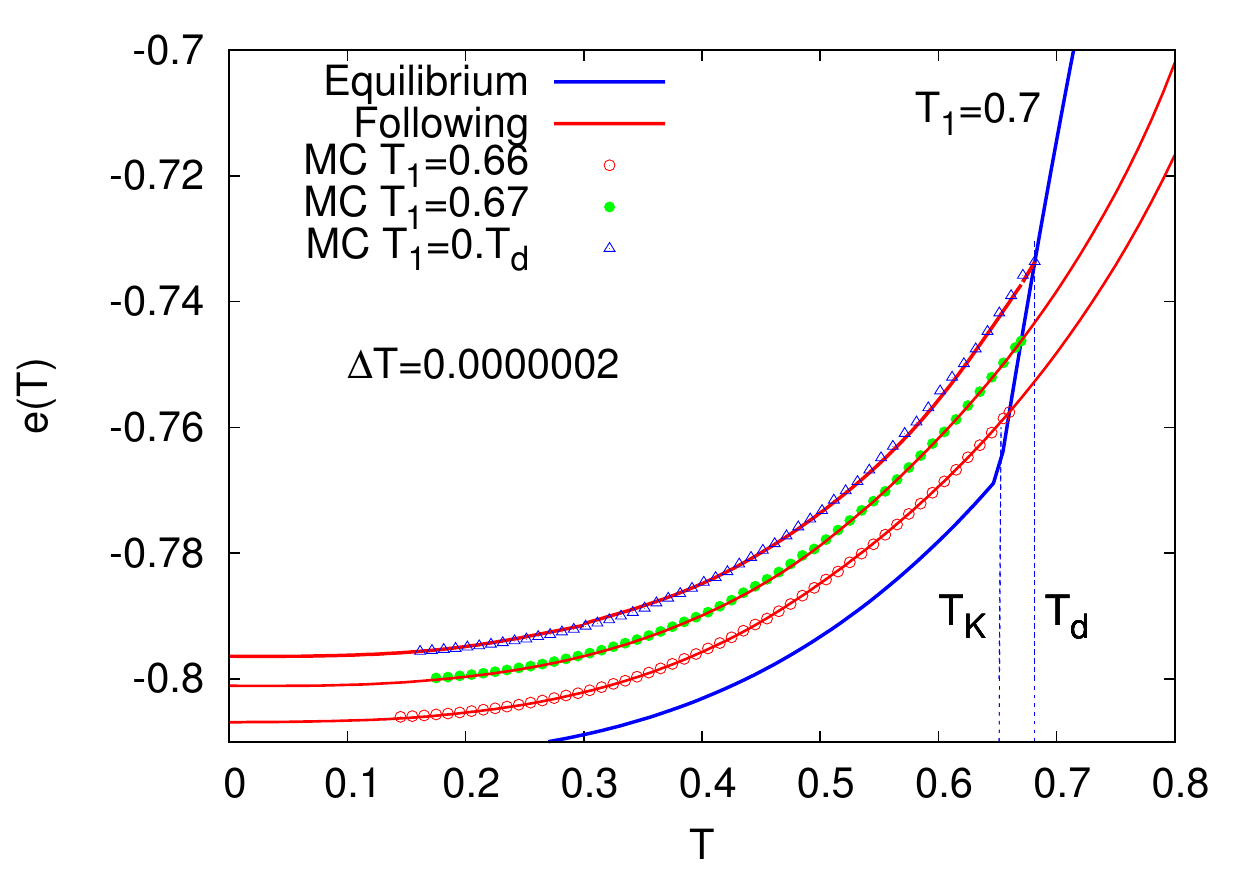}
\hspace{-0.5cm}
\includegraphics[width=0.52\textwidth]{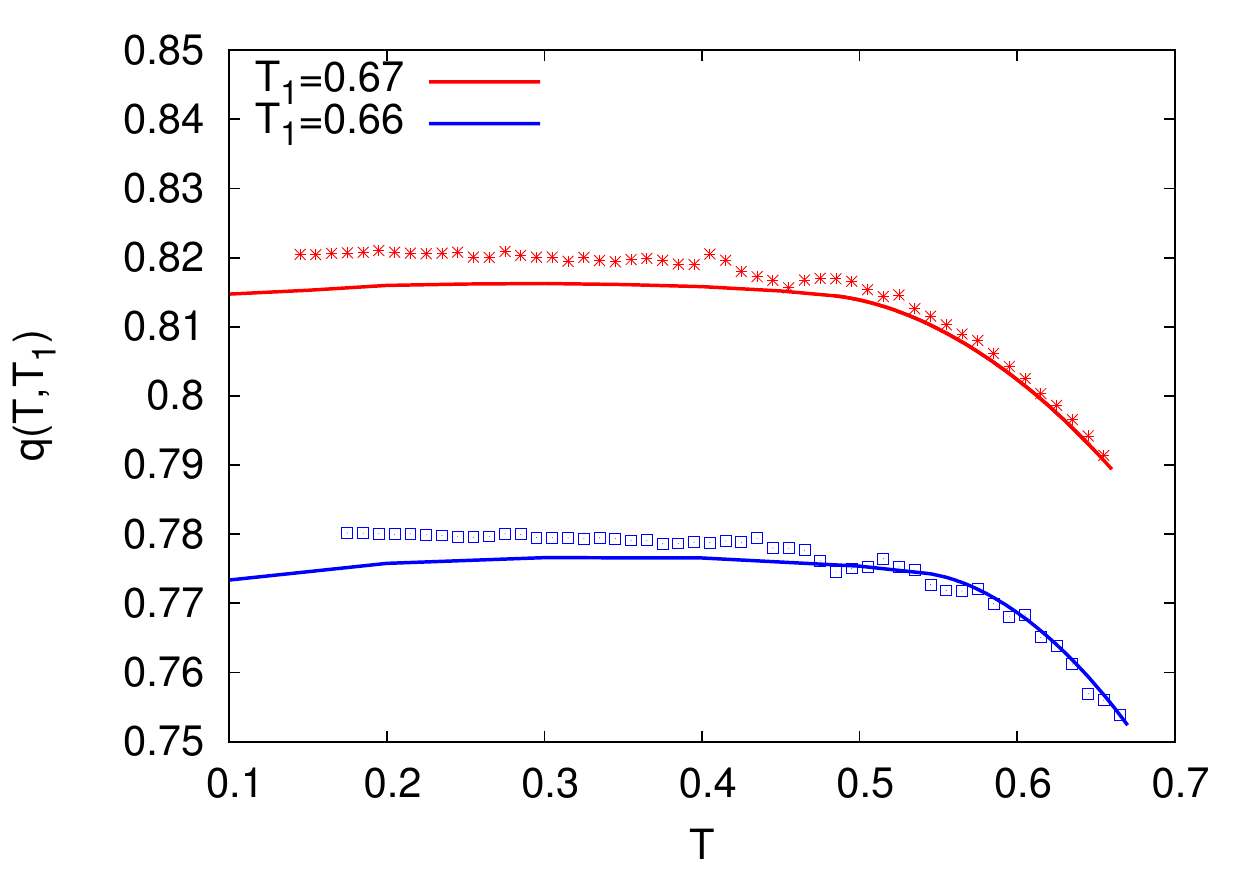}
\caption{\label{fig:2}Annealings starting from an equilibrated
  configuration at low temperature $T_1 \le T_d$ in the $3$-spin glass
  problem for $N=50000$. Left: The energy of annealings from three
  different starting temperatures $T_1=T_d, 0.67, 0.66$ with slow
  cooling rate is shown together with the equilibrium solution of the
  model (in blue) and the following state prediction (in red),
  demonstrating a perfect agreement.  Right: overlap with the starting
  configuration at $T_1=0.67,0.66$ as a function of the
  temperature. The approximated nature of the analytical solution (in
  full line) appears here as it predicts a non monotonous behavior
  that is not seen in the Monte-Carlo simulation. Indeed, the
  following state computation is unstable towards more complicated
  symmetry breaking for $T<0.55$ ($T_1=0.67$) and for $T<0.49$
  ($T_1=0.66$).}
\end{figure}

This can also be clearly seen from annealing from temperature below
$T_d$ (see Fig.~\ref{fig:2}) where the agreement is again perfect with
the following state computations. Still, the approximative nature of
the computation should be clear from deviation in the overlap (left
figure) which display non physical behavior in the following state
computation. The point is that this computation is done in a 1RSB
formalism, in a region where a more complicated ansatz should be used
(see also \cite{sun2012following}). In fact, when the state is
unstable, one should in principle performs a more involved computation
and follows the pseudo-dynamics idea
of\cite{KrzakalaKurchan07,franz2012quasi}. This is, however, more
complicated, and we shall not attempt to discuss these points here.

\section{What about quenches?}

Finally, we also performed direct quenches from a high temperature to
a small one. The situation here is more complicated as there is no theory
equivalent to the following state computation yet (but see
\cite{franz2012quasi}). 

\begin{figure}[!ht]
\hspace{-0.2cm}
\includegraphics[width=0.52\textwidth]{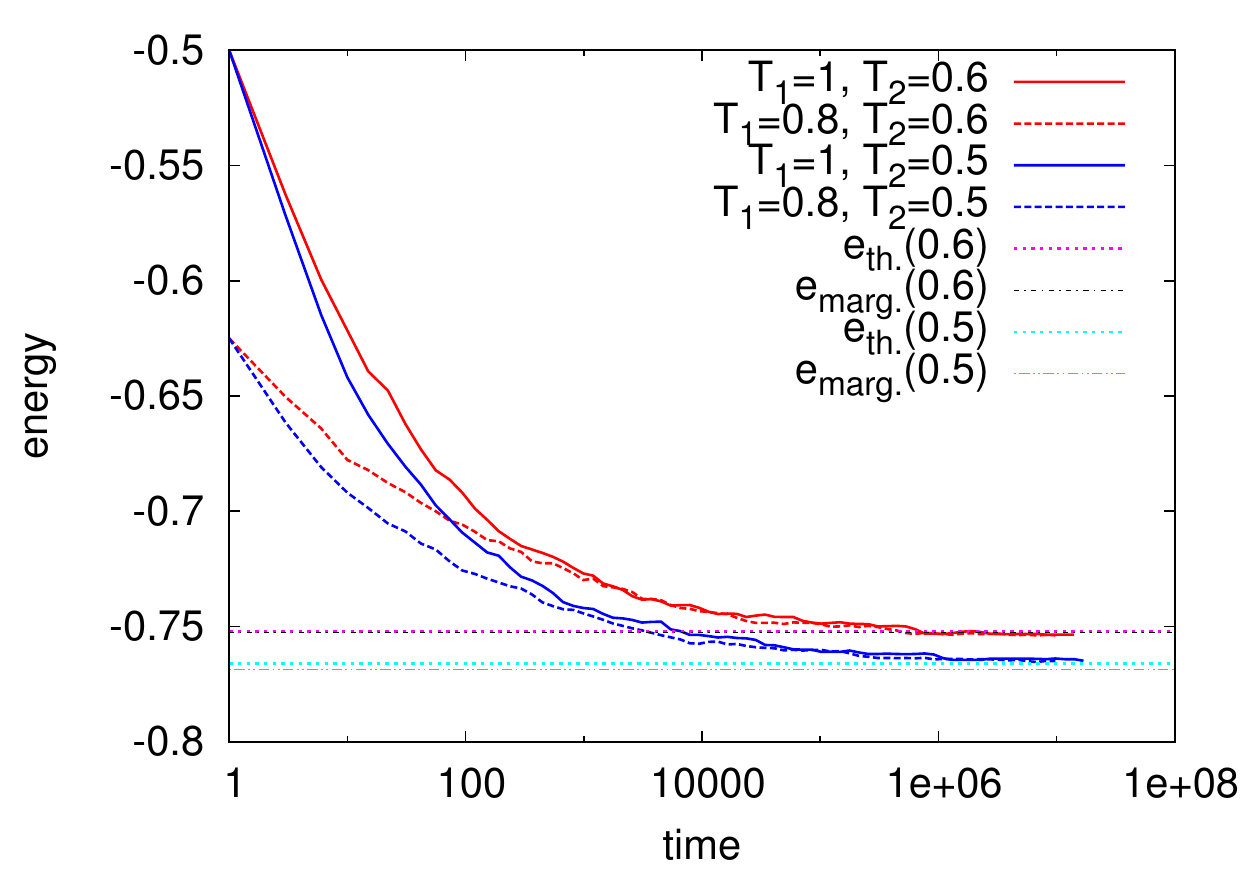}
\hspace{-0.5cm}
\includegraphics[width=0.52\textwidth]{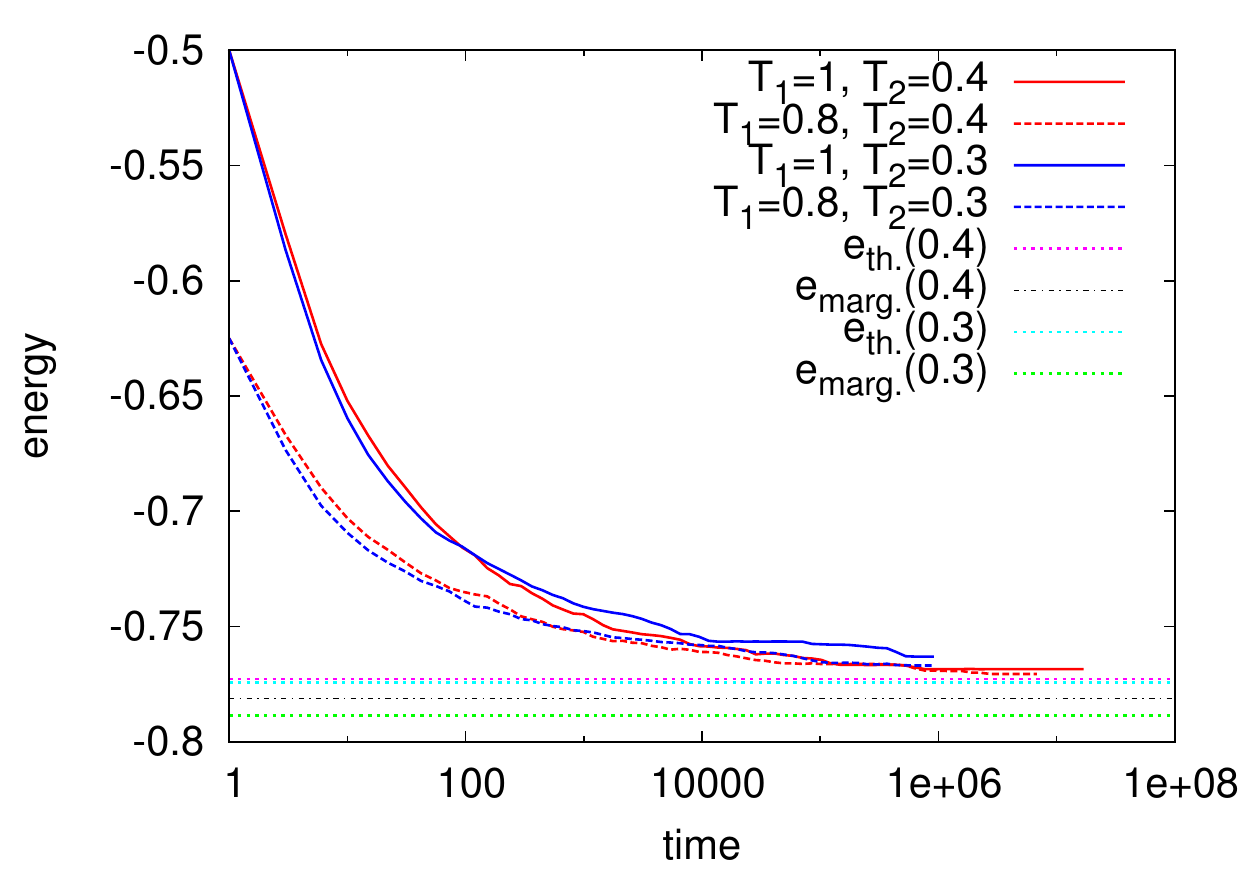}
\caption{\label{fig:3}Quenches starting from equilibrated configuration at
  temperature $T_1>T_d$ to $T_2<T_d$ for $N=20000$. Left: quenches to $T_2=0.6$ and
  $T_2=0.5$, Right:quenches to $T_2=0.4$ and  $T_2=0.3$. Although an
  extrapolation is difficult, it seems ---especially in the right
  figure--- that the asymptotic energies are larger than the marginal
  threshold states prediction. However, perhaps surprisingly, they are close to the standard
  unstable threshold states, as also noticed in \cite{MontanariRicci04}.}
\end{figure}

However, the simulation seems to indicate that the quenches converge
to higher energies than the annealing. In fact, they are surprisingly
close to the thresholds state (but, again surprisingly, higher from
the marginal ones). Although it is difficult to draw firm conclusion
from these data, the role of the marginal thresholds in the dynamics
is not apparent. Also, it is not clear that the starting temperature
is relevant, which also seem contradictory to the idea that quenches
are different to annealing in the long runs. The most recent idea,
proposed by Rizzo \cite{rizzo2013replica}, proposed that (up to
logarithmic corrections) quenches will tend to a new threshold
energy. This is a promising idea that await a numerical
confirmation. Surprisingly (given the long lasting interest in spin
glasses) we are aware of very few simulations of this problem, and a
deep, systematic investigation, would be welcome.

\section{Conclusion and perspective}
\label{conclusion}

The conclusion of this short proceeding are two folds: 1) we wanted to
illustrate how good the prediction of the following state are (and how
good the isocomplexity approximation can be in some case) and 2)
nothing seems to happens at energies connected to thresholds or
marginal threshold states. We hope that this presentation of the data
will convince the community that the bridge between statics and
dynamics is not going through thresholds states consideration, but
rather through following states considerations.

For quenches, instead, the situations is unclear and more work on the
subject is welcome, in particular to test the proposition of
\cite{rizzo2013replica}.

We hope that this short presentation will help to clarify the
situation and will help to motivate further works in this direction,
perhaps following the idea of
pseudo-dynamics\cite{KrzakalaKurchan07,franz2012quasi}.

\section*{Acknowledgments}
This work has been supported in part by the ERC under the European
Union's 7th Framework Programme Grant Agreement 307087-SPARCS, by the
Grant DySpaN of "Triangle de la Physique", and by the National Science
Foundation under Grant No. PHYS-1066293 and the hospitality of the
Aspen Center for Physics.

\section*{References}

\bibliographystyle{iopart-num}
\bibliography{biblio_tex,old}

\end{document}